\documentclass{aa}
\usepackage[varg]{txfonts}
\usepackage{graphicx}
\usepackage{amstext}
\usepackage{color}
\usepackage{natbib}
\usepackage{calc}

\begin{document}

\title{Observations of asymmetric velocity fields and gas cooling \\ in the  NGC 4636 galaxy group X-ray halo}

\author{Jussi Ahoranta\inst{1}
\and Alexis Finoguenov\inst{1}
\and Ciro Pinto\inst{2}
\and Jeremy Sanders\inst{3}
\and Jelle Kaastra\inst{4,5}
\and Jelle de Plaa\inst{4}
\and Andrew Fabian\inst{2}}

\institute{Department of Physics, University of Helsinki, FI-00014 Helsinki, Finland
\and Institute of Astronomy, Madingley Road, CB3 0HA Cambridge, United Kingdom
\and Max-Planck-Institut für extraterrestrische Physik, Giessenbachstrasse, 85748 Garching, Germany
\and SRON Netherlands Institute for Space Research, Sorbonnelaan 2, 3584 CA Utrecht, The Netherlands
\and Leiden Observatory, University of Leiden, P.O. Box 9513, NL-2300 RA Leiden, The Netherlands}

\date{Received 7 October 2015/
Accepted 14th June 2016}

\abstract
{}
{This study aims to probe the thermodynamic properties of the hot intragroup medium (IGM) plasma in the core regions of the NGC 4636 galaxy group by detailed measurements of several emission lines and their relative intensities. 
}
{We analyzed deep XMM-\emph{Newton} Reflection Grating Spectrometer (RGS) data in five adjacent spectral regions in the central parts of the NGC 4636 galaxy group. We examined the suppression of the \ion{Fe}{xvii} resonance line (15.01 Å) as compared to the forbidden lines of the same ion (17.05 Å and 17.10 Å). The presence and radial dependence of the cooling flow was investigated through spectral modeling. Parallel analysis with deep Chandra Advances CCD Imaging Spectrometer (ACIS) data was conducted to gain additional information about the thermodynamical properties of the IGM.}
{The plasma at the group center to the north shows efficient \ion{Fe}{xvii} ion resonant scattering, yielding $(I_{\lambda17.05}+I_{\lambda17.10})/I_{\lambda15.01}$ line ratios up to 2.9$\pm$0.4, corresponding $\text{to about}{}$ twice the predicted line ratio. In contrast, no resonant scattering was detected at the south side. The regions featuring resonant scattering coincide with those embodying large amounts of cool ($kT\lesssim0.4\mathrm{keV}$) gas phases, and the spectral imprints of cooling gas with a total mass deposition rate of $\sim0.8$ M$_{\sun}$ yr$^{-1}$ within the examined region of $2.4\arcmin\times 5.0\arcmin$.}
{We interpret the results as possible evidence of asymmetric turbulence distribution in the NGC 4636 IGM: Turbulence dominates the gas dynamics to the south, while collective gas motions characterize the dynamics to the north. X-ray images show imprints of energetic AGN at both sides, yet we find evidence of turbulence heating at the south and gas cooling at the north of the core. We infer that the observed asymmetry may be the result of the specific observation angle to the source, or arise from the turbulence
driven by core sloshing at south side.}


\titlerunning{Observations of asymmetric turbulence fields}
\maketitle

\section{Introduction}{\label{intro}}

The intergalactic space within galaxy clusters and groups is filled with diffuse highly ionized plasma with virial temperatures ranging from sub-keV to $\sim 10$ keVs. X-ray spectra of such intergalactic plasmas consist of a thermal bremsstrahlung continuum and line emission of several high-Z elements, whose intensities and line widths carry information about the thermodynamical state and composition of the emitting gas. In this paper we focus on detailed spectroscopic studies of properties of the X-ray halo surrounding a nearby giant elliptical galaxy, NGC 4636, the dominant galaxy of a group located at the outer parts of the Virgo galaxy cluster.

The NGC 4636 galaxy group is approximately 15 Mpc distant and its apparent X-ray luminosity is one of the brightest of all
group luminosities. Dynamically the group consists of a concentrated dark matter halo (\citealt{schuberth06}) filled with baryonic gas, with a temperature of $kT\sim0.5$ keV in the cool dense core region and $kT\sim0.8$ keV in the outer parts (e.g., \citealt{loewenstein02,sullivan05,finoguenov06,johnson09}).  
It is plausible that the core region of the intragroup medium
(IGM) halo contains multiphase plasma, as discussed in \cite{sullivan05},
for example.

The NGC 4636 X-ray halo has a complex morphology. The most prominent structures are two arms extending toward opposite sides of the galaxy, both of which connect to hot ellipsoidal bubble-shaped X-ray cavities several kpc from the central galaxy (see Fig. \ref{figure:morphology}). As scrutinized in \cite{baldi09}, for
instance, these structures are related to a large, ancient AGN outburst into the surrounding IGM gas. The outburst took place $\sim2-3\cdot10^6$ years ago (\citealt{baldi09}), releasing a total energy of $\sim6\cdot10^{56}$ ergs into the IGM (\citealt{jones02}) through relativistic jets. Observations of the radio jets directed toward the X-ray cavities were published in the multiwavelength study by \cite{giacintucci11}, indicating that the interaction between the central black hole (BH) and IGM is currently weaker. Such feedbacks can heat the IGM through turbulence (e.g. \citealt{zhuravleva14}), sound wave (\citealt{fabian12}), or weak shock heating (\citealt{randall15}). In addition to the AGN feedbacks, often quoted mechanisms capable of driving turbulence and heating the IGM include galactic merging and core sloshing (\citealt{aschasibar06}). 

Theory and simulations suggest that IGM plasma heating occurs mainly through the dissipation of turbulent kinetic energy into thermal plasma energy (see, e.g., \citealt{peterson06,hillel14} and references therein). Turbulence can also mix multiphase gas, reducing the temperature phase distribution, hence slowing down the processes leading to cooling flows, as discussed in \cite{banerjee14},
for example. 
It is currently not fully understood to what extent AGN heating counterbalances radiative cooling in individual cool-core systems such as the NGC 4636. Nevertheless, if the cool gas phases are formed by cooling from the hot phase and the central supermassive black hole remains in a quiescent state for a sufficiently long period of time, large cooling flows should develop that might trigger a new cycle of high-energy AGN outbursts (see, e.g., \citealt{jones02,werner13}). As the core region of the NGC 4636 group has been exposed to a large AGN outburst in the relatively recent past, it is a suitable source for studying the consequences for present gas thermodynamics in different regions of the X-ray halo.
The radiative cooling time in the core is considerably shorter than the Hubble time (e.g., \citealt{mathews03,chen07}), and therefore cooling flows toward the center of gravity may be present as a consequence of decreased thermal pressure of radiatively cooling IGM gas. Predictions of the traditional cooling flow model (\citealt{fabian94}) suggest total mass deposition rates of approximately $\dot{M}\sim1-2$ M$_{\sun}$ yr$^{-1}$ for the NGC 4636 group (e.g., \citealt{bertin95,chen07}). 

The NGC 4636 group's core IGM temperature gives rise to efficient emission of the \ion{Fe}{xvii} ion line complex (\citealt{doron02}). The XMM-\emph{Newton} RGS instrument with the designed wavelength band of $\sim 6-38$ Å ($\sim 0.3-2.5$ keV) is currently the most powerful observational tool to resolve and study these lines. For typical IGM conditions the \ion{Fe}{xvii} forbidden lines are optically thin, but the resonance line can become optically thick in dense regions with low turbulent velocities. Consequently, detailed measurements of the \ion{Fe}{xvii} emission can give important information on the IGM properties. Resonant scattering in galaxy clusters was first discovered by \citet{gilfanov87}, while observations of \ion{Fe}{xvii} resonant scattering in the NGC 4636 group core region has previously been published by \citet{xu02} and \citet{werner09}.

We investigated the \ion{Fe}{xvii} resonant scattering by comparing the intensities of the unresolvable blend of the forbidden line doublet at 17.05 and 17.10 Å to that of the resonance line (15.01 Å). Observations of the $(I_{\lambda17.05}+I_{\lambda17.10})/I_{\lambda15.01}$ ratio yields information on the gas dynamics, since the emission line optical thicknesses depend on both the ion column density and gas velocity distributions. 
Therefore spatial observations of resonant scattering may be used in identifying regions of turbulent and collective gas motions. However, since the RGS measurement technique relies on the use of grating elements, the spatial information is only obtained in the cross-dispersion direction at the instrument's focal plane, limiting its applicability in such measurements. In this study we have taken full advantage of this spatial dimension by analyzing long-exposure data of NGC 4636 and using five adjacent spectral regions extracted in the RGS cross-dispersion direction. We also fit the spectra with a cooling flow model to study the magnitude and radial characteristics of the flow. In addition, independent thermal analyses were conducted using the Chandra ACIS instrument to gain relevant 2D data to interpret the results of the RGS analyses. 
The paper is organized as follows: In Sect. \ref{data} we present the observational data and in Sect. \ref{preparation} the data preparation methods and analyses. In Sect. \ref{discussions} we discuss the results, and we conclude in Sect. \ref{concl}.

\section{Observational data}
\begin{figure}
\includegraphics[natheight=570pt, natwidth=800pt,width=\linewidth]{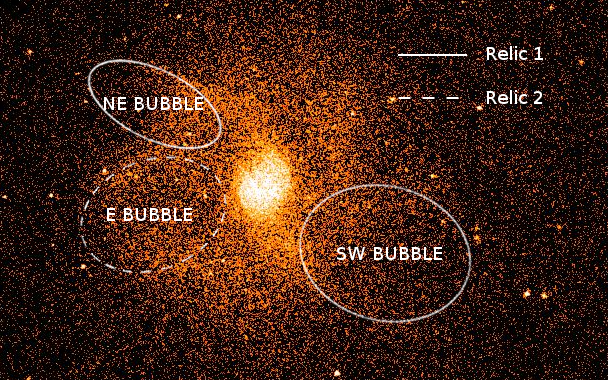}
\caption{Chadra ACIS surface brightness image of the NGC 4636 X-ray halo with bubble features contoured and labeled. Relic 2 is thought to have formed as a result of a different AGN outburst that predates the outburst that formed Relic 1, which indicates
a cyclicity of the AGN activity in NGC 4636.}
\label{figure:morphology}
\end{figure}

\begin{figure}[h]
\includegraphics[natheight=570pt, natwidth=800pt,width=\linewidth]{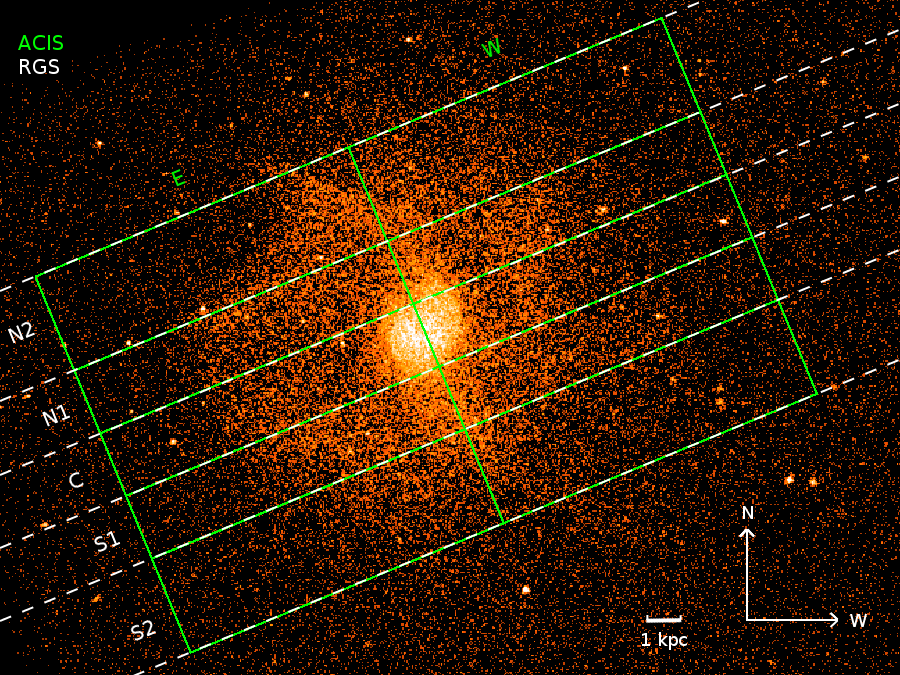}
\caption{ACIS (green rectangles) and RGS (white dashed lines) extraction region contours projected on top of the ACIS exposure of NGC 4636 together with the labels of the naming convention used in this paper. The RGS regions in the dispersion direction are unlimited.}
\label{figure:regions}
\end{figure}

{\label{data}}
In this study we used deep XMM-\emph{Newton} RGS exposures included in the CHEmical Enrichment Rgs cluster Sample (CHEERS) (\citealt{plaa15}) and deep Chandra ACIS data. The observation identification numbers and exposure times are given in Table \ref{table:2}. The CHEERS dataset includes several other RGS exposures of the NGC 4636 group (Obs. IDs 111190101/201/501), totaling an exposure time
longer by $\sim1.5$ times than in the observation used in this analysis. Observation 111190701 was chosen as its position angle and photon statistics match our aims to individually examine the emission from the different features present in the NGC 4636 X-ray halo (see Figs. \ref{figure:morphology} and \ref{figure:regions}). Using a single exposure also avoids complications that later
arise from combining data of spectra from different spatial regions.

\begin{table}
\caption{Information of the data samples used in the RGS analysis and 2T ACIS analysis.}
\label{table:2}
\begin{tabular}{l c c}
\hline\hline
Instrument & XMM-\emph{Newton} RGS & Chandra ACIS-I  \\
\hline 
Observation ID & 0111190701 & 4415 \\
Exposure time (s) & 64354 & 75347   \\
Clean time (s) & 58466 & 74363\\
\hline
\end{tabular}
\tablefoot{
ivo://ADS/Sa.CXO\#obs/04415
}
\end{table}

%
\section{Data preparation and analysis}{\label{preparation}}
%
\subsection{XMM-\emph{Newton} RGS instrument}{\label{RGS}}

The RGS data reduction was performed with the XMM-\emph{Newton} Science Analysis Software (XMM SAS) version 13.5.0 \footnote{http://xmm.esac.esa.int/sas/}. We created the calibration files with the SAS task \emph{cifbuild} and used the \emph{rgsproc} pipeline to create the data products. We extracted the light curves for the two RGS instruments with the \emph{evselect} procedure to find the correct flux thresholds for neglecting the time intervals of elevated soft protons that
are reflected into the RGS focal planes from the telescope mirrors. The \emph{tabgtigen} procedure was used to create good time interval (GTI) files and the \emph{rgsproc} in data filtering. 

The NGC 4636 group extends over the field of view of the RGS instrument, which prohibits the use of outer regions of RGS data in background subtraction. We used a modeled background spectrum instead, which is based on blank field observations and scaled by the count rate in RGS CCD9, in which only low emission from the source is expected. 

XMM-\emph{Newton} EPIC MOS data were used to generate the surface brightness profile projections along the RGS dispersion-direction, separately for each RGS extraction region. These profiles were used in corrections for instrumental line broadening by the spectral analysis software, as discussed later in this section. 
The MOS data reduction was performed similarly to that of the RGS reduction; the main differences being the different SAS procedures that were used to create the data-products (\emph{emproc}). The cumulative surface brightness projections were generated through \emph{rgsvprof}.

To enable the investigation of the source spectra as a function of the cross-dispersion direction, we implemented an extraction scheme consisting of five adjacent spectral regions centered on the core region. In angular units, we extracted a $0.4\arcmin$ section centered on the peak of the measured surface brightness profile (region C), an adjacent $0.4\arcmin$ wide region on the north side to C (region N1) and a $\sim0.6\arcmin$ wide region adjacent to N1 (N2). Corresponding regions were extracted symmetrically from the south side of region C (S1, S2). This segmentation scheme was found to be the best compromise between the spectral quality and our exploration interests. The approximate extraction region projections relative to the ACIS surface brightness map are shown in Fig. \ref{figure:regions} with dashed lines.

The regions were defined by generating extraction masks in the \emph{rgsproc} pipeline by the use of the SAS parameters \emph{xpsfincl} and \emph{xpsfexcl}. These parameters define the fraction of point-source events to be included within the extraction mask and the fraction of point-source events to be excluded in the background mask, respectively. The \emph{srcxdsp} parameter was used to define the cross-dispersion positions of the masks.
We focused on getting the regions to correspond to the quoted values as precisely as possible within the wavelength band including the \ion{Fe}{xvii} emission lines. The detailed angular information of the extraction regions is given in Table \ref{table:RGS}. 

\begin{table}
\caption{Angular coordinates and spectral region widths at the $14.5-17.5$ Å wavelength band. The quoted values are associated with an uncertainty of $\sim\pm0.02\arcmin$.}
\label{table:RGS}
\begin{tabular}{l c c }
\hline\hline
Reg. & RGS 1+2 (arcmin)  & width (arcmin) \\
\hline 
N2 &  $-1.21...-0.64$ & 0.57 \\
N1 &  $-0.63...-0.21$ & 0.42 \\
C  &  $-0.20...+0.20$ & 0.40 \\
S1 &  $+0.20...+0.61$ & 0.41 \\
S2 &  $+0.58...+1.19$ & 0.61 \\
\hline
\end{tabular}
\end{table}

Incoming photons from different spatial angles along the dispersion direction project photons shifted on the RGS focal plane $\lambda$-axis according to the dispersion angle $\Delta \theta$. This causes broadening of the integrated spectral line shapes, which in first-order spectra is described by
\begin{equation}
\Delta \lambda = 0.138 \cdot \Delta\theta \ Å,
\end{equation}
where $\Delta\theta$ is given in the arcmins. The cumulative line-broadening profile files based on the MOS surface brightness profiles were used to correct for this broadening.

Spectral fitting of the RGS data was performed with SPEX v. 2.04.00 (\citealt{kaastra96})\footnote{http://www.sron.nl/SPEX}. For the plasma models we used the most recently updated proto-solar abundances of \cite{lodders09} and the  ionization balance data
of \cite{bryans09}. We fit the spectra in the $8-28\mbox{ Å}$  wavelength range and corrected for redshift (0.0037, \citealt{chen07}) and for galactic absorption ($n_{H}=2.07\cdot10^{24}\mbox{m}^{-2}$)\footnote{http://www.swift.ac.uk/analysis/nhtot/}.
As NGC 4636 IGM most likely contains multiphase plasma, we chose to fit the data with a two-temperature collisional Ionization equilibrium (CIE) model, which enabled modeling the emission of the main CIE components at each spectrum. This way, more accurate fitting of the thermal continuum was achieved for detailed measurements of emission line intensities. In the cooling flow measurements, we replaced the second CIE component by a cooling flow model and used the CIE model mainly to fit the element abundances.

The resonant-scattering spectral fits were performed in two stages. In the first stage we neglected the spectral region containing the \ion{Fe}{xvii} lines susceptible to resonant scattering ($13.8-15.5\mbox{ Å}$) and found the best fit for the thermal continuum component. This was done to eliminate the effects that the resonant scattering may otherwise induce in our initial fits, given that the line spectra are dominated by the set of strong Fe lines. We fit the two CIE temperatures, their normalizations ($\int{n_e n_H dV}$), and SPEX parameter \emph{dlam}, which fits the linear spectral shift in the RGS wavelength axis.
We used the generated MOS surface brightness profiles in the SPEX spatial broadening model (\emph{lpro}) and fit the \emph{lpro} broadening parameter \emph{s} separately for both CIE components. This was done since it is likely that the spatial distribution of emitting ions for a given CIE component differs somewhat from that predicted by the \emph{lpro} model, which is based on the total intensity variations of the surface brightness profile. We also fit the abundances of the elements in the spectrum, that
is, N, O, Ne, Mg, and Fe. Except for Fe, the abundances were coupled between the two CIE components. The Fe abundances were uncoupled in the spectra where photon statistics allowed to do so (S1, C, and N1), since this enabled more accurate fitting of the Fe emission lines in the spectra. Lack of photon statistics prevented us from doing this for other elements. After finding the best initial fit, we froze the temperatures and Fe abundances in the model.

In the next fitting stage we retained the excluded spectral region and removed the \ion{Fe}{xvii} ion from our model. We used the SPEX delta line model to regenerate the set of \ion{Fe}{xvii} emission lines, which enabled us to deviate from the line intensity dependences and hence accurately measure the $(I_{\lambda17.05}+I_{\lambda17.10})/I_{\lambda15.01}$ line ratios. These lines were multiplied with the \emph{lpro} model so that they corresponded the spectral line shapes of our best-fit model. The lines were generated for the 2p-3d transitions at 15.015 Å and 15.262 Å and the 2s-3d lines at 16.77 Å and (the blend) at 17.077 Å, which are the main \ion{Fe}{xvii} emission lines present and resolvable in the spectra. We then fit the delta line normalization parameters and obtained the $(I_{\lambda17.05}+I_{\lambda17.10})/I_{\lambda15.01}$ line ratios of the delta line normalizations. We used C-statistics and 1-$\sigma$ errors in parameter error calculations. 

The results of the RGS analysis are presented in the upper panel of Table \ref{table:1} and are discussed in Sects. \ref{rs} (resonant scattering) and \ref{cf} (cooling flow).

\subsection{Chandra ACIS instrument}{\label{ACIS}}

\subsubsection{Two CIE modeling of rectangular extraction regions}{\label{2T}}

The CIAO v. 4.6 scripts and CALDB 4.6.3 \footnote{http://cxc.harvard.edu/ciao/} calibration libraries were used in the data reduction for the 2T spectral analysis. We performed the data processing using the standard codes available for extended sources. The data were reprocessed with the \emph{repro} script, which creates the necessary bad-pixel, event, and PHA files. We then generated the GTI files with the \emph{dmgti} to exclude solar flares from the data. For background subtraction we used the appropriate blank-sky files from CALDB, since the source emission is extended over the field of view of the instrument. We used CIAO \emph{wavdetect} to remove point sources and converted the data into SPEX format with \emph{specextract} and the SPEX task \emph{trafo}. 

ACIS spectra were prepared for five $4\arcmin$ long extraction regions (E+W, see Fig. \ref{figure:regions}), in a way that their position angle matched the RGS observation and the widths in RGS cross-dispersion direction those of the RGS regions. We also extracted two symmetrical spectral regions from each of these five regions, the eastern (hereafter E) and western (W) halves.

Similarly to the RGS analysis, the two CIE component model was used when fitting to find the dominate thermal components in each spectral region in the 0.5 to 2.0 keV energy band. If the best fit for a given region converged into a single temperature model, a refit with a single-temperature model was used. The free parameters in fitting were the temperatures (range $0-1.5$ keV), normalizations of the CIE component(s), coupled O, Ne, Mg, and Fe abundances, and $n_H$ of the galactic absorption model. The reason for fitting the $n_H$ in the thermal analysis followed from findings of additional absorption in the NGC 4636 X-ray halo outside the center, particularly in the most distorted regions of the X-ray halo (\citealt{mushotzky94,ohto03}). A correction for redshift of 0.0037 was applied. We used C-statistics because of the low total count numbers per extraction region and used 1-$\sigma$ errors in error calculations.

The results of the ACIS temperature measurements are presented in the lower panel of Table \ref{table:1} and are discussed in Sect. \ref{dis_2T}.

\subsubsection{ACIS spectral maps}{\label{maps}}

The spectral maps were created by fitting spectra extracted from Chandra observation identifiers 323 (ACIS-S), 3926 (ACIS-I) and 4415 (ACIS-I). The datasets were reprocessed using acis\_process\_events, enabling VFAINT mode for observations 3926 and 4415. To exclude flares, we made light curves in 200 s bins using CCD 7 for observation 323 and CCDs 0, 1, and 2 for the ACIS-I observations, clipping time periods where the count rate was outside $2.5\sigma$ using a sigma-clipping algorithm. After filtering, the total exposure time was 189 ks. Datasets 3926 and 4415 were reprojected to match the coordinate system of dataset 323.  We obtained standard blank-sky background event files for each observation, reprocessed them as above, applied VFAINT mode as appropriate, reprojected them using the appropriate observation aspect file to match the observation, and then reprojected the event file to match the coordinate system of dataset 323. The exposure time of the background event files was adjusted to match the count rate in the 9 to 12 keV band. The exposure times of the background event files were further reduced by discarding events, so that the ratio of each background exposure to the total background exposure time was the same as the ratio of the respective foreground exposure to the total foreground exposure. Point sources were identified using the CIAO task \emph{wavdetect} and were excluded from analysis.

Spatial bins were created using the contour binning algorithm (\citealt{sanders06b}) to have a signal-to-noise ratio of at
least 20 ($\sim 400$ counts), using a geometric constraint factor of 2. Spectra were extracted from each foreground and background event file from these regions. Responses and ancillary responses were created for each bin and observation, weighting by the number of counts in the 0.5 to 7 keV band in each response region. The foreground and background spectra for the two ACIS-I observations were added and the responses averaged. 

The spectra were fit in the 0.5 to 7 keV band using XSPEC (\citealt{arnaud96}), fitting the ACIS-I and -S data simultaneously, by minimizing the C-statistic. The spectra were fit with the APEC thermal model v. 2.0.2 (\citealt{smith01}), applying the PHABS absorption model of Balucinska-Church \& McCammon (\citeyear{balucinskachurch92}) with an equivalent hydrogen column of $1.9 \times 10^{24}$ m$^{-2}$. We assumed the relative solar abundances of Anders \& Grevesse (\citeyear{anders89}) and fixed the abundance to 0.59 Z$_{\odot}$. This made the (pressure and entropy) spectral maps more stable, but had only a minor effect on the temperature map. As a result of using a different spectral analysis software  than in the 2T fits, different physical models and model parameters were adopted. For this reason, an additional temperature map was generated with SPEX, which enabled more transparent comparison with the results of the 2T fits. The SPEX T-map the data were fit with the 1 CIE model, using abundances by \citet{lodders09} fixed to solar value, and $n_{H}=2.07\cdot10^{24}\mbox{m}^{-2}$. 

\begin{figure}
\includegraphics[natheight=2538pt, natwidth=1391pt,width=\linewidth]{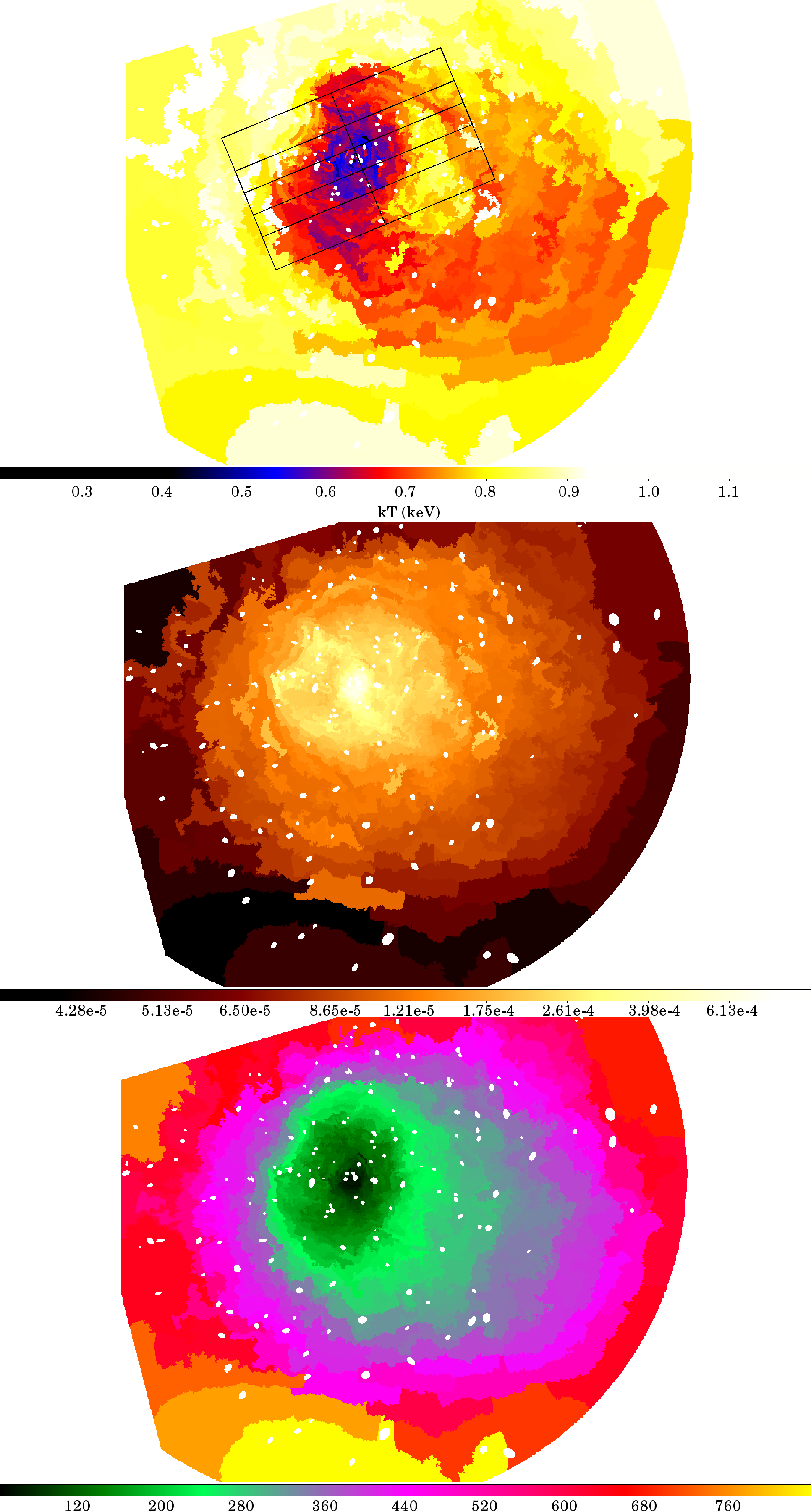}
\caption{Top panel: The projected temperature map of the NGC 4636 central regions with rectangular ACIS extraction region contours added for a reference. Middle and bottom panels: Pseudo-pressure map (above) and pseudo-entropy map (below) plotted with arbitrary units. The T-map is based on SPEX CIE modeling, whereas the pressure and entropy maps where prepared using the APEC model.
 }
\label{figure:tmap}
\end{figure}

The spectral maps are presented in Fig. \ref{figure:tmap} and discussed in Sect. \ref{dis_maps} We present the SPEX 1T temperature map, the XSPEC pseudo-pressure map ($n_{\mathrm{pse}}^{1/2}\cdot kT$), and a pseudo-entropy map ($n_{\mathrm{pse}}^{-1/3}\cdot kT$), where the $n_{\mathrm{pse}}$ denotes for the XSPEC normalization ($10^{-14}/(4\pi[D_A(1+z)]^2) \int{n_e n_H dV}$) per square-arcsec. In Sect. \ref{comparison} we compare the results of the 2T and T-map analyses.

%
\section{Discussion}{\label{discussions}}

\subsection{Chandra measurements}{\label{chandra}}

\subsubsection{Spectral maps}{\label{dis_maps}
NGC 4636 is a well-studied galaxy group (e.g., \citealt{jones02,schuberth06,baldi09,johnson09,giacintucci11}). Using a 1T model and azimuthally averaged spectral data, the NGC 4636 IGM's radial temperature distribution can be described by power laws. However, the 2D temperature map (the upper panel of Fig. \ref{figure:tmap}) shows that in the context of this study, the temperature phase distribution is insufficiently described by spherical symmetry. 

The T-map shows an extension of cold gas toward the south, shaped similarly to the examples of core sloshing (see, e.g., \citealt{markevitch07}). The pseudo-pressure and -entropy maps attest for this assertion, showing low-entropy gas along the tail of the cold gas (lower panels in Fig. \ref{figure:tmap}), possibly due to the ram-pressure stripping of gas from the core region. The spectral maps therefore suggest a current motion of the NGC 4636 core towards the northeast in the sky plane projection. 

Since the NGC 4636 elliptical is the dominant galaxy of a cool-core group, we interpret the core sloshing as the result of gravitational disturbance in the core region, for instance,
caused by a merging event. The spectral maps do not show a spiral shape, which is often associated with the core sloshing, which could be due to either an early evolutionary stage of the sloshing phenomenon (see, e.g., \citealt{zuhone15}), to the specific observation angle (see figures in \citealt{zuhone15b}), or both. However, based on the spectral maps, we would expect sloshing motions of the gas to be present at least in the C and S side spectral regions. Simulations suggest that sloshing-induced gas flows can drive turbulence into the gas, although its effect on total turbulence fields at the central regions of galaxy clusters or
groups is generally considered to be smaller than that of AGN feedbacks (see, e.g., \citealt{vazza12}).

\subsubsection{Fitting with two-phase model}{\label{dis_2T}}

The Chandra ACIS data were fitted with a 2T model to describe the dominant temperature phases at each RGS extraction region. The analyses were performed for 4$\arcmin$ wide extraction regions matching the RGS cross-dispersion direction extraction scheme (regions E+W, see the definitions in Fig. \ref{figure:regions}). In addition, we divided each of these region into eastern and western sections, which enabled us to perform the spectral analyses for regions enclosing the main morphological features of the X-ray halo separately. The results of these fits are given in Table \ref{table:1} and are plotted in Fig. \ref{figure:CIE}. 

In the table we present the 1T fit results for each region and 2T results for those regions where the modeling yielded two different CIE components. We found that making use of 2T modeling improves C-statistics in the central regions and at the N side (except for sector N1E), and that the most significant improvements were yielded for regions where cold ($kT<0.4$) gas components were detected.

\begin{figure}
\resizebox{\hsize}{!}{\includegraphics{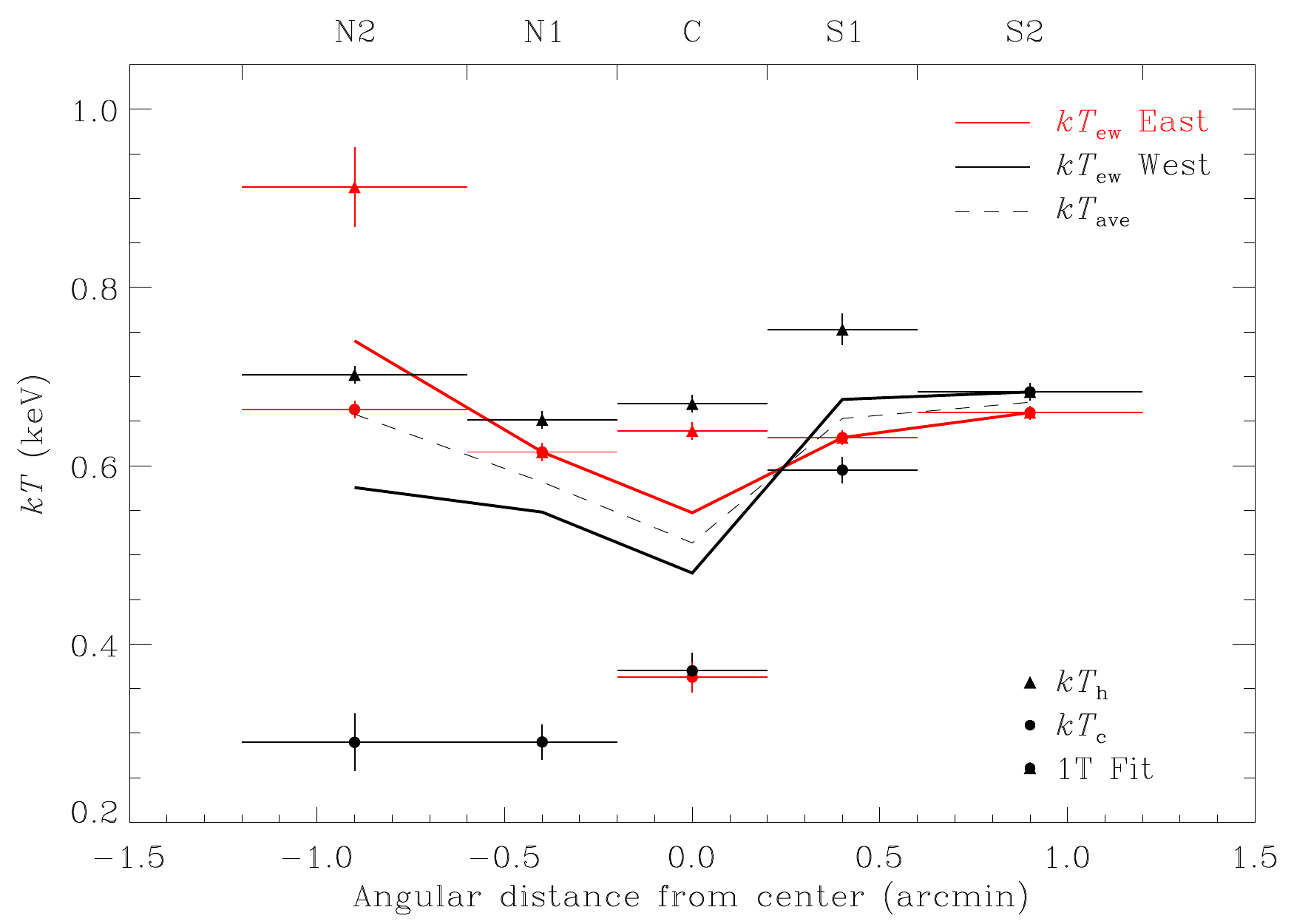}}
\caption{Best-fit values for temperatures of the two-component CIE models for ACIS regions plotted together with emission-weighted temperatures (solid lines) and their average (dashed line). The triangles and circles mark the hot and cold components, respectively.
}
\label{figure:CIE}
\end{figure}

\begin{figure*}
\resizebox{\hsize}{!}{\includegraphics{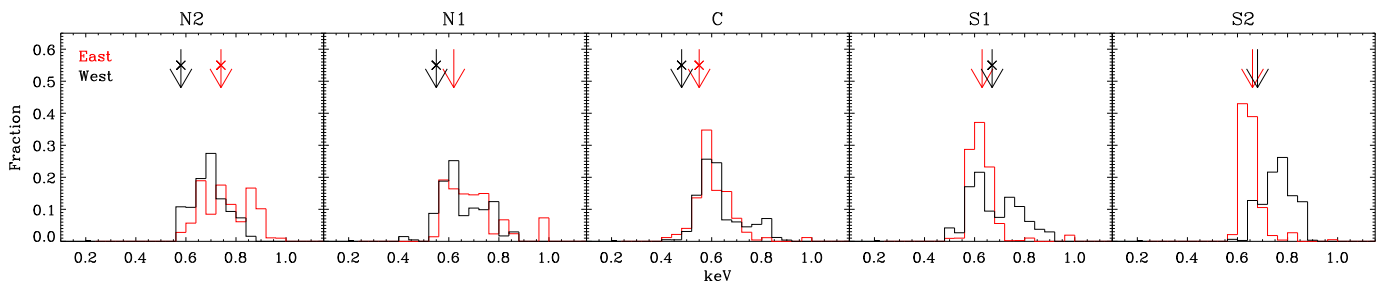}}
\caption{
Comparison of the two-temperature analyses. The histograms show the distributions of the 1T fits of the temperature map within each rectangular extraction region, weighted by the SPEX normalizations per arcsec. The arrows indicate the best-fit values obtained in the 2T fitting cycle. The crossed arrows show the best-fit values of the emission-weighted temperatures of the two CIE component models, while the arrows without the crosses mark the values of the fits reduced in the 1T model. 
}
\label{figure:temp_dist}
\end{figure*}

We found evenly distributed gas phases with $kT\sim0.6-0.7$ keV, which constitutes the majority of the IGM gas within the examined volume. Hot, $kT\gtrsim0.75$ keV gas phases were found coinciding spatially with the X-ray cavities of the AGN feedback Relic 1 (as defined in Fig. \ref{figure:morphology}). The hot and cold phases found in these regions agree with the temperature jumps related to the X-ray cavities published by \citet{baldi09}. In addition to the hot phases, the fits yielded cold $kT\la0.4$ keV CIE components in regions NW(1-2), CW, and CE. These cold gas components were most prominent in the central regions, where they constituted $\text{about}$ half of the total emission measure in the spectra. No cool components were significantly detected at the extraction regions enclosing AGN outburst Relic 1, nor at the regions of ongoing jet activity published by \cite{giacintucci11} (regions NE1-2 and SW1-2). This implies that the detected cold gas phases may
  originate from radiative cooling in quiescence regions, and that this cooling is more efficient in the N than in the S side.

In Fig. \ref{figure:CIE} we plot the emission-measure-weighted temperatures of the spectral regions with solid lines and their average with dashed lines in the direction of the RGS cross-dispersion axis. The temperature gradients between the N and S side are similar, although we found large scatter of temperature phases from the center to north, whereas on the S side the temperature seems to be distributed more evenly. This may imply more efficient turbulence mixing of the gas phases on the S side. While the two-phase fitting suggests multiphase gas to be present in the less disturbed regions at the north side (NW1-2), no multiphase gas was significantly detected in the quiet sectors at the south side (SE1-2).

\subsubsection{Comparison of the results of 2T fits and temperature map}{\label{comparison}}

To compare the results of the two-temperature analyses, we present the distribution of the individual 1T fits of the SPEX T-map and the results of the 2T fitting for each rectangular extraction region in Fig. \ref{figure:temp_dist}. The values of the T-map were weighted by the SPEX normalizations per square arcsecs to make them more easily comparable with the results obtained in the two-phase fitting round.

The 2T fits yield temperatures lower by $\sim0.1$ keV than the T-map fits in most of the regions, which likely expresses the involvement of systematic errors that are due to differences in modeling. The comparison also suggests that a wider temperature distribution within a region causes a larger disparity between the results of the two analyses. Otherwise the two data sets match reasonably well and indicate the same findings of the temperature structure of the source; the temperatures are more scattered in the regions N2-C than in the S side regions. 

Both modeling sets hence reassert the impression that the regions N2-C consist of multiphase gas, while the south side is described well with only two main temperature components: A widely spread $0.6-0.7$ keV phase, and the hot gas related to the SW bubble.

\begin{table*}
\caption{List of the fitted and derived values for the NGC 4636 spectra. The subscripts \mbox{h} and \mbox{c} denote hot and cold components, and $Y_\mathrm{c}/Y_\mathrm{h}$ for the SPEX emission measure ratio of these components. The kT$_\mathrm{ew}$ are the SPEX emission-measure-weighted temperatures. All the abundances are reported relative to H, in solar units.}
\label{table:1}
\centering
\begin{tabular}{l c c c c c c c c c c}
\hline\hline
&&&& \textbf{RGS} \\
\hline

Parameter & N2 && N1 && C && S1 && S2 \\
\hline 
$(I_{\lambda17.05}+I_{\lambda17.10})/I_{\lambda15.01}$ & 2.03$\,\pm\,$0.32 && 2.90$\,\pm\,$0.40 && 2.08$\,\pm\,$0.25 && 1.20$\,\pm\,$0.16 && 1.35$\,\pm\,$0.25 \\

$I_{\lambda15.01}^{\mathrm{obs}}/I_{\lambda15.01}^{\mathrm{pred}}$ & $0.64\pm0.08$ && $0.45\pm0.05$ && $0.62\pm0.06$ && $1.08\pm0.13$ && $0.96\pm0.14$ \\

Fe$_\mathrm{c}$ & - && $0.66_{-0.05}^{+0.14}$ && $0.52_{-0.02}^{+0.07}$ && $0.61\,\pm\,0.08$ && - \\
Fe$_\mathrm{h}$ & - && $0.96\,\pm\,0.06$ && $0.95\,\pm\,0.10$ && $0.96\,\pm\,0.10$ && - \\
Fe$_\mathrm{}$ & $0.84\,\pm\,0.03$ && - && - && - && $0.55\,\pm\,0.03$ \\

O &  $0.92\,\pm\,0.09$ && $0.58\,\pm\,0.06$ && $0.45\,\pm\,0.04$ && $0.62\,\pm\,0.07$ &&  $0.48\,\pm\,0.07$ \\

C-stat/d.o.f. & $752/639$ && $761/637$ && $761/637$ && $753/637$ && $820/639$ \\

\hline 
$\dot{M}$ (M$_{\sun}\mbox{yr}^{-1}$) & 0.08$\,\pm\,0.01$ && 0.25$\,\pm\,0.01$ && 0.35$\,\pm\,0.01$ && 0.04$\,\pm\,0.01$ && 0.10$\,\pm\,0.01$ \\
C-stat/d.o.f. & $654/550 $ && $657/550$ && $658/550$ && $690/550$ && $706/550$ \\

\hline

&&&& \textbf{ACIS} \\

\hline
\textbf{EAST:}\\

\textbf{2T:}\\
kT$_\mathrm{h}$  (keV)  & $0.91\,\pm\,0.05$ && -                  && $0.64\,\pm\,0.01$   && -                    && -                    \\
kT$_\mathrm{c}$ (keV)   & $0.66\,\pm\,0.01$ && -                  && $0.36\,\pm\,0.02$   && -                    && -                    \\
$Y_\mathrm{c}/Y_\mathrm{h}$     & $2.24$    && -                  && $0.50$               && -                   && -                    \\
kT$_\mathrm{ew}$ (keV)  & $0.74$            && -                  && $0.55$               && -                   && -                    \\
C-stat/d.o.f.           & $174/123$         && -                  && $285/123$    && -                   && -    \\
\hline
\textbf{1T:}\\
kT (keV)                & $0.69\,\pm\,0.01$ &&  $0.62\,\pm\,0.01$ &&   $0.61\,\pm\,0.01$ &&  $0.63\,\pm\,0.01$  && $0.66\,\pm\,0.01$         \\
C-stat/d.o.f.           & $175/125$         && $183/125$          &&     $296/125$        && $229/125$           && $210/125$            \\
\hline
F$_\mathrm{h}$ (10$^{-13}$erg s$^{-1}$ cm$^{-2}$) & $1.11$ && $3.88$ && $5.90$ && $4.49$ && $3.48$ \\
F$_\mathrm{c}$ (10$^{-13}$erg s$^{-1}$ cm$^{-2}$) & $2.52$  && - && $1.88$ && - && - \\
\hline 

\textbf{WEST:}\\

\textbf{2T:}\\
kT$_\mathrm{h}$  (keV)  & $0.70\,\pm\,0.01$ && $0.65\,\pm\,0.01$  && $0.67\,\pm\,0.01$    && $0.75\,\pm\,0.02$   && -                    \\
kT$_\mathrm{c}$ (keV)   & $0.29\,\pm\,0.03$ && $0.29\,\pm\,0.02$  && $0.37\,\pm\,0.02$    && $0.60\,\pm\,0.02$   && -                    \\
$Y_\mathrm{c}/Y_\mathrm{h}$     & $0.44$    && $0.40$             && $1.73$               && $0.99$              && -                    \\              
kT$_\mathrm{ew}$ (keV)  & $0.58$            && $0.55$             && $0.48$           && $0.67$              && -                    \\
C-stat/d.o.f.           & $198/123$       && $177/123$    && $281/123$        && $197/123$           && -                    \\
\hline
\textbf{1T:}\\
kT (keV)                & $0.69\,\pm\,0.01$ &&  $0.65\,\pm\,0.01$ &&   $0.61\,\pm\,0.01$ &&  $0.66\,\pm\,0.01$    && $0.68\,\pm\,0.01$    \\
C-stat/d.o.f.           & $204/125$         &&  $223/125$         &&     $292/125$        &&     $197/125$       && $184/125$            \\
\hline
F$_\mathrm{h}$ (10$^{-13}$erg s$^{-1}$ cm$^{-2}$) & $3.36$ && $4.29$ && $4.14$ && $1.88$ && $3.42$ \\
F$_\mathrm{c}$ (10$^{-13}$erg s$^{-1}$ cm$^{-2}$) & $0.60$  && $0.46$ && $2.73$ && $1.98$&& - \\

\hline
E+W:\\
kT (keV) & 0.68$\,\pm\,0.01$ && 0.62$\,\pm\,0.01$ && 0.60$\,\pm\,0.01$ && 0.64$\,\pm\,0.01$ && 0.66$\,\pm\,0.01$\\
C-stat/d.o.f.           & $245/75$      && $286/75$             &&  $450/75$               && $278/75$             && $224/75$ \\
\hline \\

\end{tabular}
\end{table*}

\begin{figure}
\resizebox{\hsize}{!}{\includegraphics{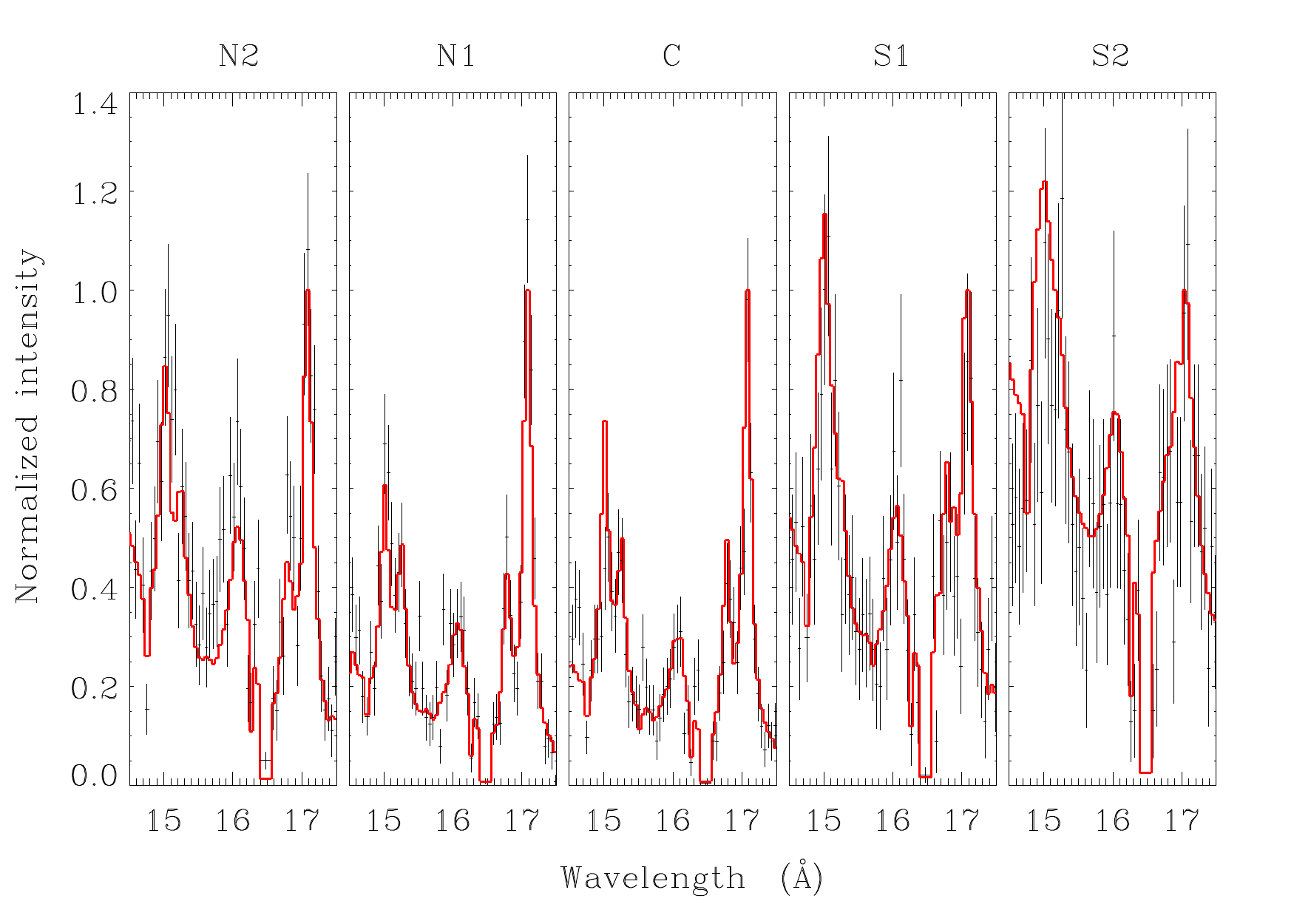}}
\caption{
RGS \ion{Fe}{xvii} emission lines as seen in each region. Each spectrum has been normalized to unity according to the best-fit model intensities of the \ion{Fe}{xvii} 17 Å lines for comparison.
}
\label{figure:Fexvii}
\end{figure}

\subsection{Resonant scattering measurements}{\label{rs}}

\subsubsection{\ion{Fe}{xvii} measurements}{\label{fe}}

The resonant scattering measurements focused on examining the $\ion{Fe}{xvii}$ line intensity ratios between the resonance line at 15.01 Å to that of the blend of the 17.05 Å and 17.10 Å emission lines. The optical thickness of a given line is proportional to the product of its ion column density and transition oscillator strength, which for the $\ion{Fe}{xvii}$ ion 15.01, 17.05, and 17.10 Å lines are about $10^{0}$, $10^{-1}$ , and $10^{-7}$, respectively (see, e.g., \citealt{deplaa12}). The line optical thickness also depends upon the gas velocity distribution in the line of sight because a photon is only absorbed if its energy closely matches the energy of the transition in the absorbing ion's frame of reference. Hence the amount of resonant scattering is smaller in the gas with broadened velocity distributions because
of the Doppler effect. 

When a photon is absorbed in the resonant scattering process, photon with the same energy will be emitted almost instantly, but to random direction. This can lead to several outcomes that affect the observations. For instance, the complex gas dynamics of the absorbing gas can broaden the resonance line with respect to optically thin lines. In high optical depths ($\tau>10$), the line energy conversions become important (\citealt{wood00}). Both of these processes lead to suppression of the resonance line intensity as compared to those of the non-resonant lines. On the other hand, in turbulent gas, the resonant scattering effect is suppressed only through small-scale turbulent motions. Moreover, the resonant scattering 
efficiency is correlated with the micro-turbulence velocity distribution with respect to the observer, being much stronger in tangential than in radial velocity components (\citealt{zhuravleva11}). Finally, if the shape of the absorbing gas volume is asymmetric, the net result of multiple scatterings will cause the resonance line flux to be enhanced in the direction of the lowest optical depth while it is reduced elsewhere (e.g., \citealt{kaastra95}). Hence observational results may depend not only on the gas density fields and dynamics, but on the particular measurement geometry as well.

Observations of the $\ion{Fe}{xvii}$ resonant scattering within the core region of the NGC 4636 was first published by \cite{xu02}. \cite{werner09}  investigated the resonant scattering in the center ($0.5\arcmin$ region) and outside the core by combining data of two $2.25\arcmin$ regions extracted from the two sides of the central slit, finding resonant scattering only in the core. We studied the dependence of the resonant scattering on the radial distance from the center with the method introduced by \cite{werner09}. These results are listed in Table \ref{table:1}. In the table we give the measured $(I_{\lambda17.05}+I_{\lambda17.10})/I_{\lambda15.01}$ ratios and quote the $I_{\lambda15.015}^{\mathrm{obs}}$/$I_{\lambda15.015}^{\mathrm{pred}}$ compared to the prediction of the three-ion collisional-radiative model published by \cite{doron02}, while adopting the $kT$ from the 1T fits of the E+W regions.

We point out that there is currently a controversy about the correct numerical values of the $(I_{\lambda17.05}+I_{\lambda17.10})/I_{\lambda15.01}$ ratios in optically thin plasma, which can result in up to 30 \% systematic uncertainty in derived values of the turbulence velocities between different models (\citealt{deplaa12}). We chose to compare our measurements to the model of \cite{doron02} because it closely agrees with the observational Fe line ratios in several IGM spectra where low Fe absorption is expected (see, e.g., the multisource study by \citealt{werner09}). However, we used the \cite{doron02} model only as a reference and did
not include it in the line ratio measurements. Hence comparisons of quoted $I_{\lambda15.015}^{\mathrm{obs}}$/$I_{\lambda15.015}^{\mathrm{pred}}$ ratios between symmetrically extracted regions are only associated with much smaller systematic errors that are induced by possible skew in the reference model Fe line ratio temperature dependency within the considered temperature range (see Fig. \ref{figure:ratiomodels}).

We detected resonant scattering in the regions N2, N1, and C. An unexpected result is that the resonant scattering is strongest several kpc off the center of the gravity, yielding $2.90\pm0.40$ in the region N1 (and $2.03\pm0.32$ in N2) compared to $2.08\pm0.25$ in the region C (see Fig. \ref{figure:resca}). Therefore the resonance line intensity in the region N1 was found to be weaker than half of the predicted value. When we compared our $I_{\lambda15.015}^{\mathrm{obs}}$/$I_{\lambda15.015}^{\mathrm{pred}}$ measurements to the simulations of the NGC 4636 resonant scattering in \cite{werner09}, we found that high turbulent velocities ($v_{turb}\gtrsim0.75c_s$) are present in the S side. In contrast, in the N side the resonance line suppression exceeds the predictions for zero turbulent velocities. This additional absorption would be best explained if the cool gas phases in the N side occurred in a concentrated instead of an evenly distributed manner. Hydrodynamical simulations of galaxy clusters suggest that gas clumping is pronounced in dynamically active systems, giving rise to concentrated clumps of cool gas with high radiative cooling rates (\citealt{nagai11}). Being in the right temperature range for efficient \ion{Fe}{xvii} emission, high resonant scattering within large overdense gas clumps could in principle enhance the Fe line ratio by a detectable amount even in spatially integrated spectral data such as those of the RGS. In this sense, the NGC 4636 IGM makes an interesting case for combining hydrodynamical simulations and observational data to investigate such cooling processes.

\begin{figure}
\resizebox{\hsize}{!}{\includegraphics{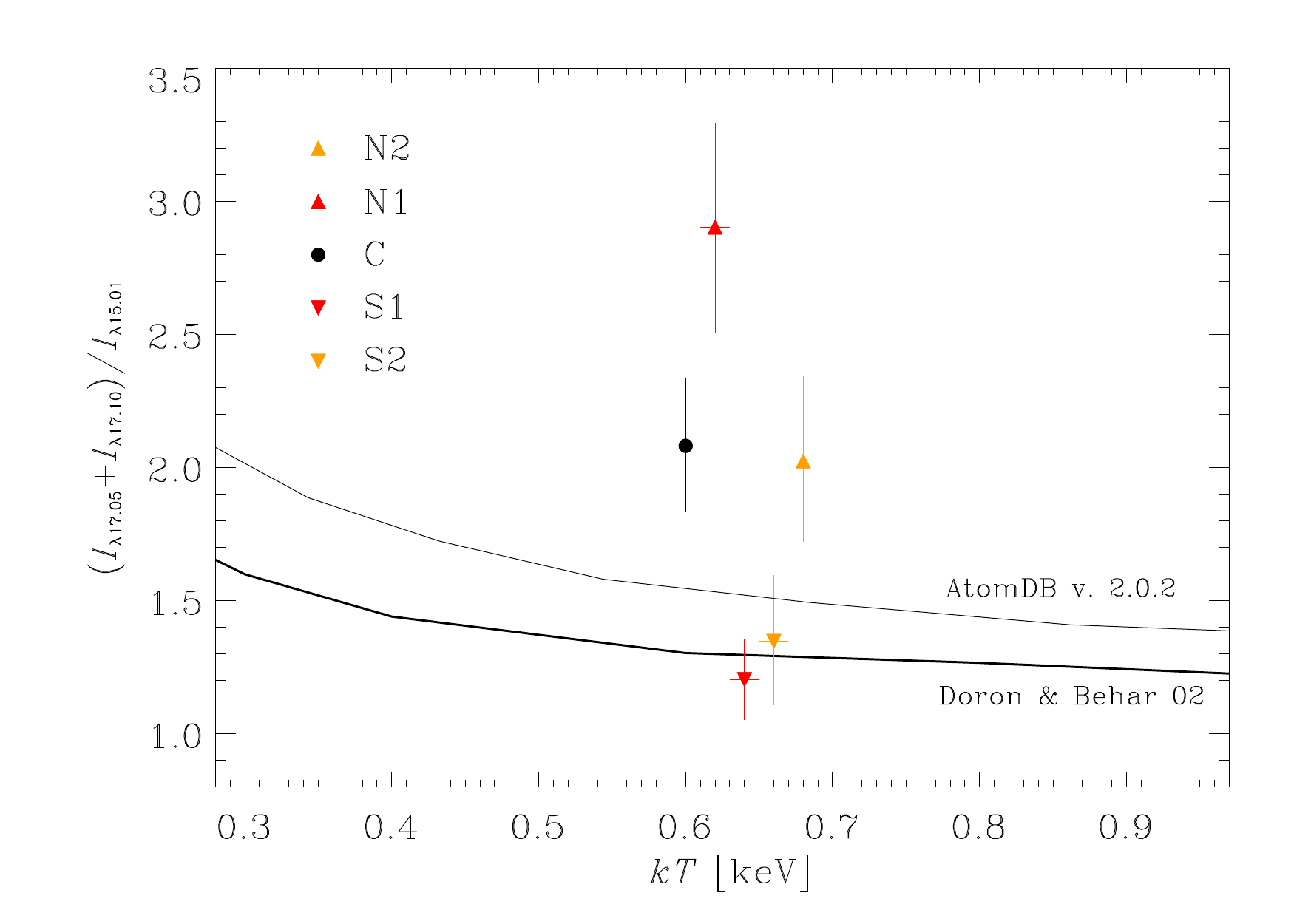}}
\caption{
Measured Fe line ratios compared with the predictions for optically thin line ratios by two different models: the AtomDB v.2.0.2 (\citealt{loch06}) and \citet{doron02}.
}
\label{figure:ratiomodels}
\end{figure}

\begin{figure}
\resizebox{\hsize}{!}{\includegraphics{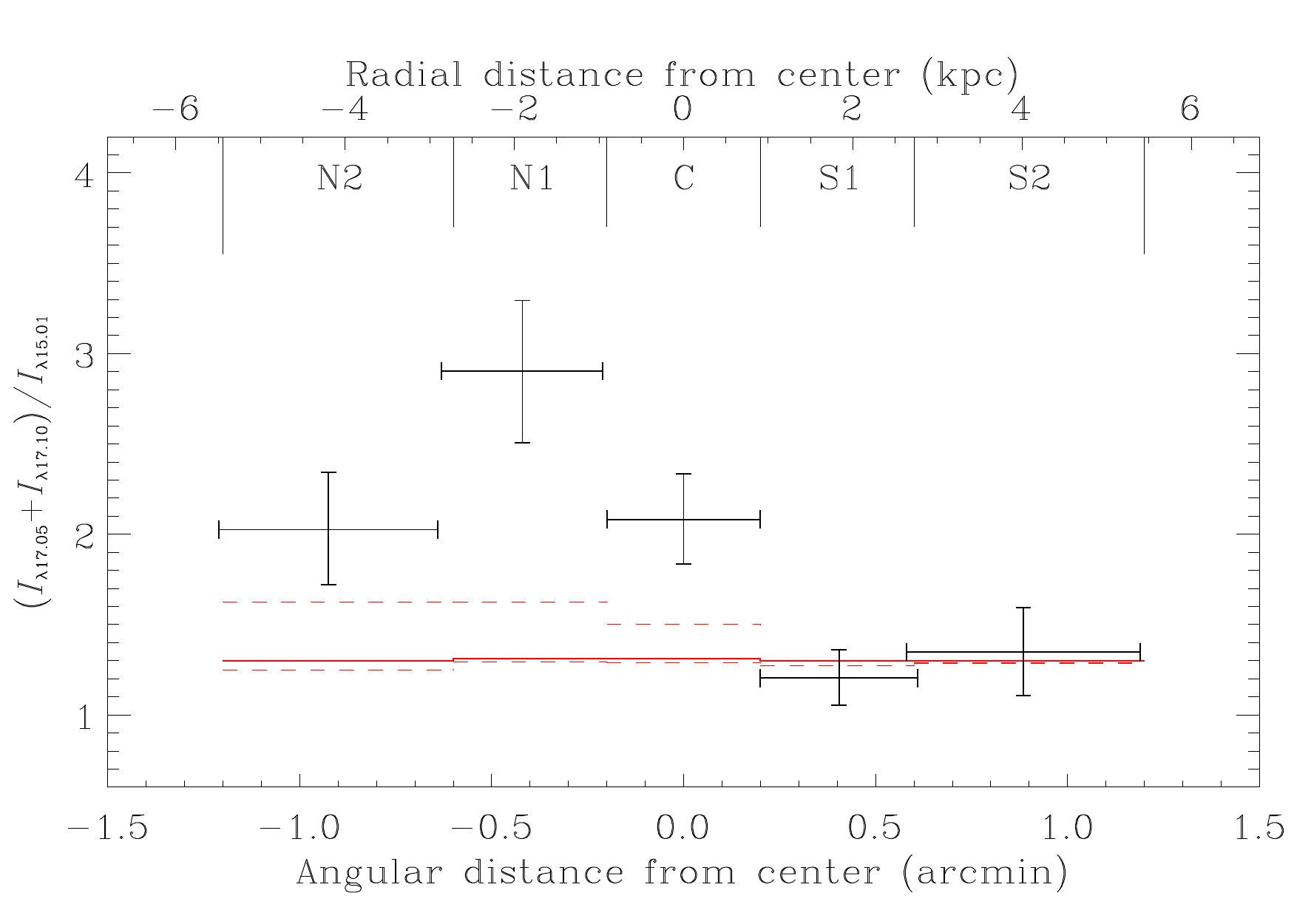}}
\resizebox{\hsize}{!}{\includegraphics{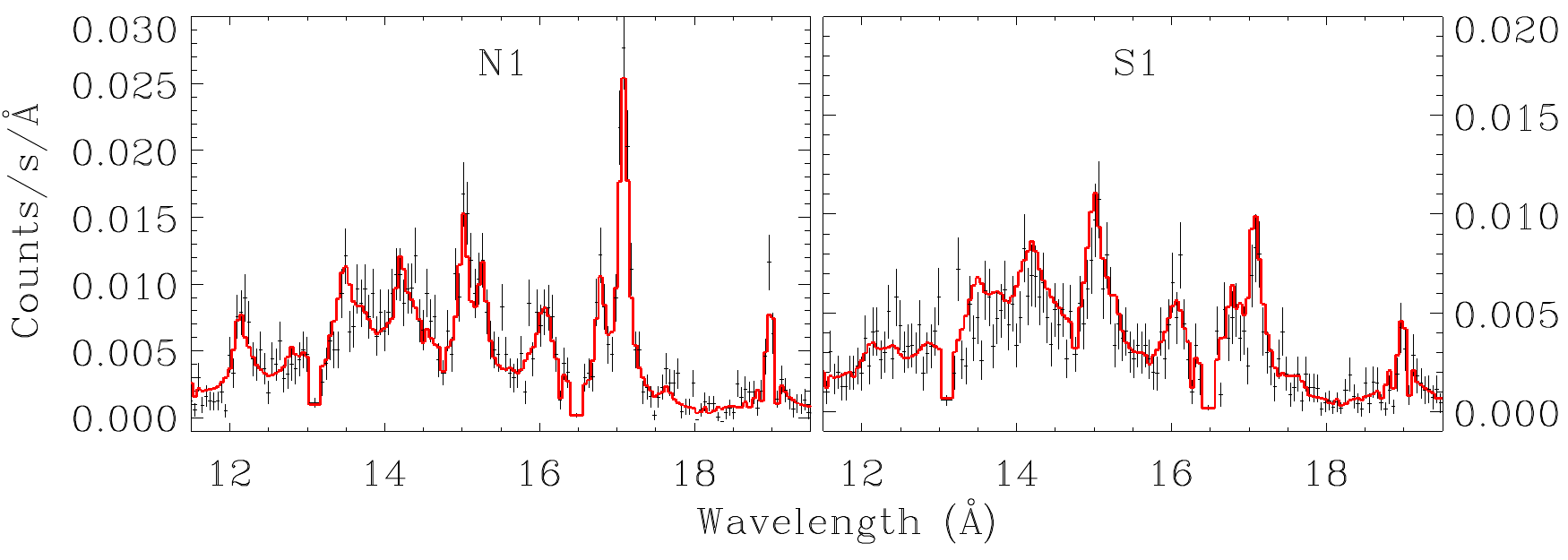}}
\caption{
Upper plot: The measured \ion{Fe}{xvii} $(I_{\lambda17.05}+I_{\lambda17.10})/I_{\lambda15.01}$ line ratios for each RGS extraction regions. The solid red line corresponds to the predicted ratio without resonant scattering for the emission-weighted temperatures for each region, whereas the dashed lines are the predicted values for the coolest (upper) and hottest (lower) phases found in each region. The distances refer to those of the RGS cross-dispersion direction in the exposure.
Below: $11.5-19.5$ Å spectra from regions N1 and S1.}
\label{figure:resca}
\end{figure}

There are also other possibilities that could explain the observed resonant scattering radial asymmetry: 1) Contributions by emission from undetected gas phases to the spectra, since the \ion{Fe}{xvii} line ratios have a temperature dependence. 2) Anomalously high Fe column densities in regions N1 and N2. 3) High optical depth of the \ion{Fe}{xvii} resonance line that are due to the absorption of the cold gas phases found in regions C to N2. 

First one of these items is ruled out through the high values obtained for the \ion{Fe}{xvii} line ratios, as is demonstrated in Fig. \ref{figure:ratiomodels}. Likewise, anomalously high Fe column densities in N1 and N2 cannot explain the observations
because the fits yield very similar Fe abundances for regions N1, C, and S1. As for the last item, we find that the asymmetric distribution of diffuse cold gas phases cannot  produce the observed resonant scattering asymmetry either. This can be seen in Fig. \ref{figure:mach}, where we plot the relation between IGM proton column density $n_p$ and the characteristic micro-turbulence velocities $V_{turb}$ for the gas phases in spectral regions S1 (red) and N1 (black).

The curves in Fig. \ref{figure:mach} were calculated by applying the relation of resonance line optical depth at the line center (see, e.g., \citealt{churazov10}): 

\begin{equation}
\label{equation:tau}
\tau=\int{\frac{\sqrt{\pi}hr_ecf}{E_0\sqrt{\frac{2kT}{Am_pc^2}(1+1.4AM^2)}}\ion{Fe}{}\delta_{\ion{}{xvii}}n_Hdl},
\end{equation}
where $r_e$ is the classical electron radius, $f$ the resonance line oscillator strength, $E_0$ the transition energy, $A$ the atomic mass of the element, $M$ the gas turbulence velocity in Mach units ($M=V_{turb}/c_{s}$), Fe the iron abundance relative to hydrogen, $\delta_{\ion{}{xvii}}$ the fraction of iron in the \ion{Fe}{xvii} ion state, and $l$ the distance in the photon propagation direction. 

To use the Eq. \ref{equation:tau} to calculate the $n_p-V_{turb}$ curves in Fig. \ref{figure:mach}, we first determined the optical depths of the \ion{Fe}{xvii} resonance line in regions N1 and S1 by using the photon escape factor method (e.g., \citealt{irons79}).  In practice, the method can be applied to lines with moderate optical depths ($\tau\lesssim5$) if the line's optically thin intensity is known. Here we used the measured $I_{\lambda15.015}^{\mathrm{obs}}$/$I_{\lambda15.015}^{\mathrm{pred}}$ ratios to determine the amount of resonance line suppression, and checked the corresponding $\tau$ from the tabulated values in the paper of \citet{kastner90} (for S1 we assumed the scattered, low 1$\sigma$ value). The benefit of this method is that it is independent of source geometry, as long as all the scattered photons can be considered to be completely lost from the line
of sight after scattering (the photon loss probability $\epsilon=1$).

We plot the $n_p-V_{turb}$ relations for two distinct conditions: for homogeneous spatial distributions of emitting and absorbing ions ($q=1$), and for completely inhomogeneous distributions ($q=0$) (Tables 1 and 5 in \citealt{kastner90}). The curves were generated by adopting the best-fit values for the $kT$ from the temperature analysis, the corresponding \ion{Fe}{xvii} ion fractions, and a value of $f=2.31$ for the resonance line oscillator strength\footnote{http://physics.nist.gov/PhysRefData/ASD/}. Along with these curves, we plot the 1- and 2$\sigma$ upper limits of $V_{turb}$ within the 0.8$\arcmin$ wide RGS slit centered on the NGC 4636 core (the blue lines, derived from \citealt{pinto15}), and independently derived $n_p$ for N1 and S1 regions (the vertical lines). These proton columns where calculated at azimuthal ($r=0.6\arcmin$) projections to the N1 and S1 regions, while integrating the column path along the inner boundaries of the N/S1 regions. This gives an estimate of the proton column densities at the gas volumes where most of the \ion{Fe}{xvii} ions are located, which makes them comparable to the plotted $n_p-V_{turb}$ curves.

Figure \ref{figure:mach} indicates considerable differences in turbulent velocities between the N1 and S1 regions. In particular, $V_{turb}$ is supersonic in the S1 region, but only a fraction of $c_{s}$ in the region N1. The high turbulent velocities at S1 could in principle be explained as due to bulk motions in the sloshing tail. The figure also indicates that depending on the true distribution of the emitting and absorbing \ion{Fe}{xvii} ions, the turbulent heating at the N1 region may not be sufficient to balance gas cooling. It is possible, however, that the $\epsilon=1$ requirement is not strictly met within the examined regions. We note that this would result in lower $V_{turb}$ values.

\begin{figure}
\resizebox{\hsize}{!}{\includegraphics{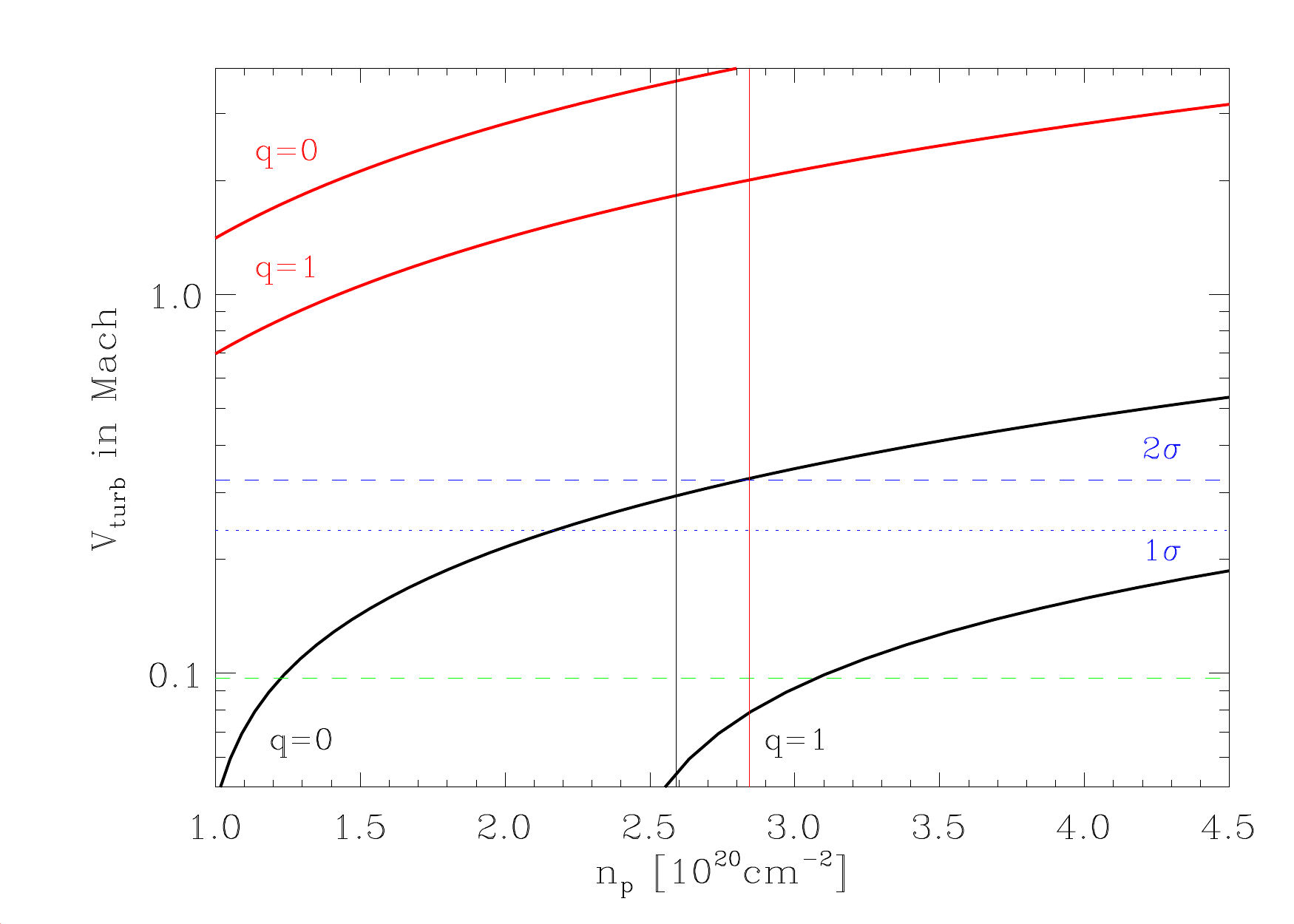}}
\caption{
Relations between IGM proton column density and characteristic micro-turbulence velocities in regions S1 (red) and N1 (black) with derived proton columns (vertical lines). $q=0$ and $q=1$ refer to inhomogeneous and homogeneous distribution of emitting and absorbing ions, respectively. The blue lines are the 1- and 2$\sigma$ upper limits for mean turbulent velocities in the NGC 4636 core regions, and the green dashed line marks the limit where turbulent heating is expected to balance gas cooling (see \citealt{pinto15}).
}
\label{figure:mach}
\end{figure}


Based on the reasons discussed above, we infer that the asymmetric resonant scattering originates from asymmetric velocity fields. The resonant scattering observations indicate that the gas velocity fields are dominated by turbulence in regions S1 and S2, whereas in regions C, N1, and N2, the turbulence is lower, implying the presence of stationary or laminar flowing gas.

We note that the $\ion{Fe}{xvii}$ emission power is highest in the temperature range of $kT\sim0.3-0.7$ keV, peaking at $kT\sim0.5$ keV. For hotter gas phases the fraction of iron in the \ion{Fe}{xvii} ion state is lower, and consequently the \ion{Fe}{xvii} line emission power drops to $\sim1/4$ of the maximum already at 0.8 keV (see the tables in \citealt{doron02}). Referring this to the thermal properties of the NGC 4636 core region gas, the hot plasma related to X-ray bubbles does not contribute significantly to the generation of the $\ion{Fe}{xvii}$ lines. For instance, the 2T fit of the RGS N1 spectrum yielded Fe$_\mathrm{h}$/Fe$_\mathrm{c}\approx1.5$, $kT_{\mathrm{h}}\approx0.86$ keV, $kT_{\mathrm{c}}\approx0.60$ keV and $Y_{\mathrm{h}}$/$Y_\mathrm{c}\approx0.5$. The Fe abundance ratio seems suspiciously high between the two phases, which may be due to degeneracies in two-phase fitting with only Fe abundances being uncoupled between the two phases, or due  to kinematically uncoupled ambient gas through which the core is moving. Nevertheless, combining these results with the information of the \ion{Fe}{xvii} line power, we find that the contribution of the hot component in the \ion{Fe}{xvii} 15.01 Å, 17.05 Å and 17.10 Å line production in the best-fit model of N1 is only $\lesssim15\%$ of that of the colder component.

\subsubsection{\ion{O}{vii} measurement}{\label{ovii}}

We extracted an additional RGS region from $-1.0\arcmin$ to $0.4\arcmin$ along the cross-dispersion axis to confirm the presence of the $\lesssim0.4$ keV gas phases found in the ACIS analysis at N1W, N2W, CW, and CE. We investigated the emission of the \ion{O}{vii} ion triplet that is emitted only at cool ($\lesssim0.4$ keV) temperatures. The \ion{O}{vii} forbidden line (22.1 Å) was observed in this spectrum with a 3.3$\sigma$ significance level (see Fig. \ref{figure:ovii}). The \ion{O}{vii} resonance line at $\lambda=21.6$ Å is strongly suppressed, indicating efficient resonant scattering within the cool phases. Furthermore, \ion{O}{vii} and \ion{Fe}{xvii} resonant scattering coincides spatially, suggesting collective gas motions of multiple gas phases occurring in the same volume. \cite{pinto14} published measurements of \ion{O}{vii} emission of the NGC 4636 group X-ray halo, observing the resonance line with 3.2$\sigma$, and the forbidden line with 3.9$\sigma$ significance. Their study was conducted using the full $3.4\arcmin \times 5.0\arcmin$ field of view of the RGS and stacked observations with comparable position angle. Comparing these results to our measurements shows that \ion{O}{vii} resonant scattering is considerably weaker outside the volume examined here.

\begin{figure}
\resizebox{\hsize}{!}{\includegraphics{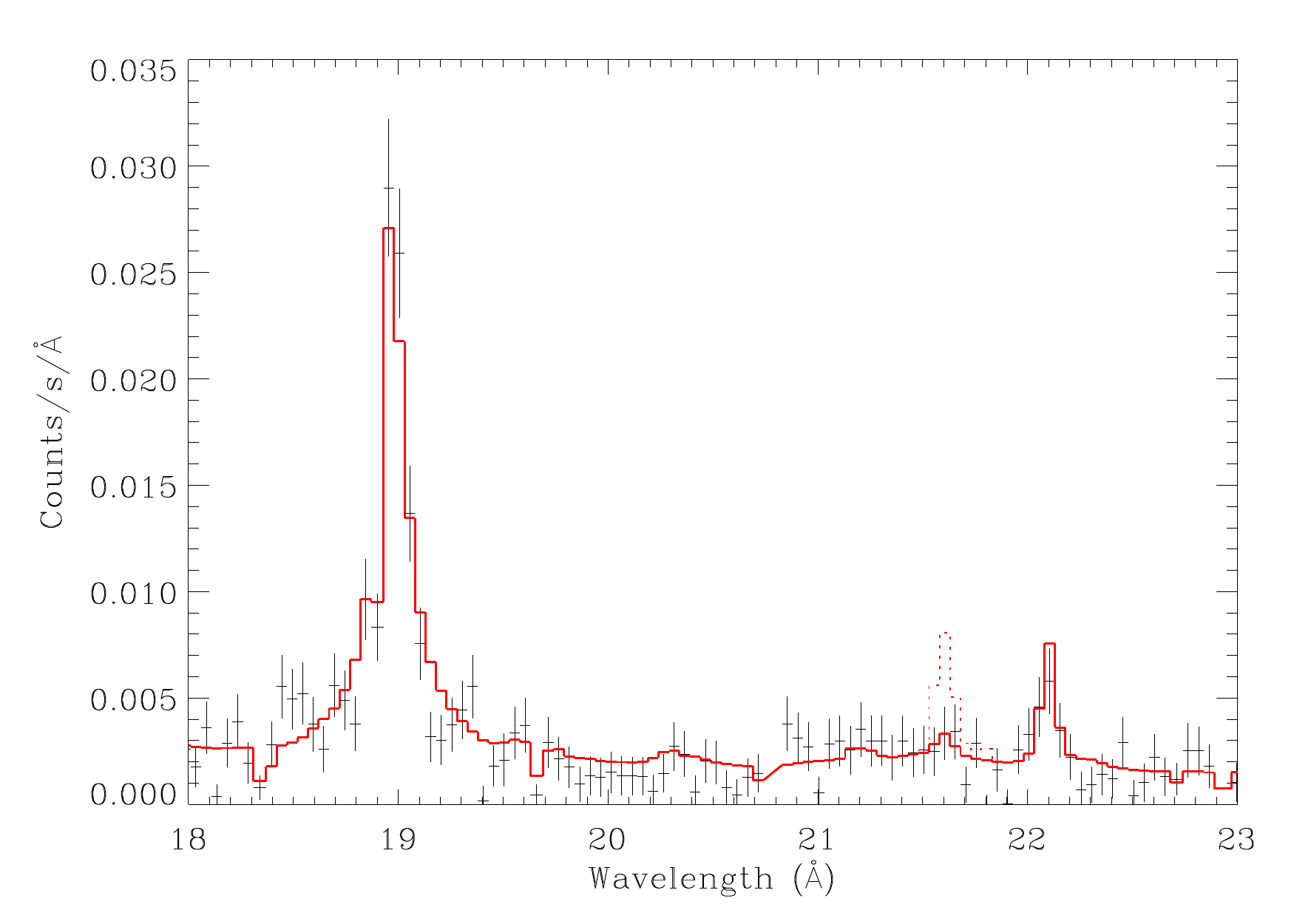}}
\caption{Zoom into \ion{O}{viii} and \ion{O}{vii} lines in a global fit of the RGS1 extraction region ($-1.0\arcmin$,$0.4\arcmin$). The dashed line shows the predicted \ion{O}{vii} triplet shape as normalized to the best-fit model forbidden line intensity, revealing strong suppression of the 21.6 Å resonance line.}
\label{figure:ovii}
\end{figure}

\begin{figure}
\resizebox{\hsize}{!}{\includegraphics{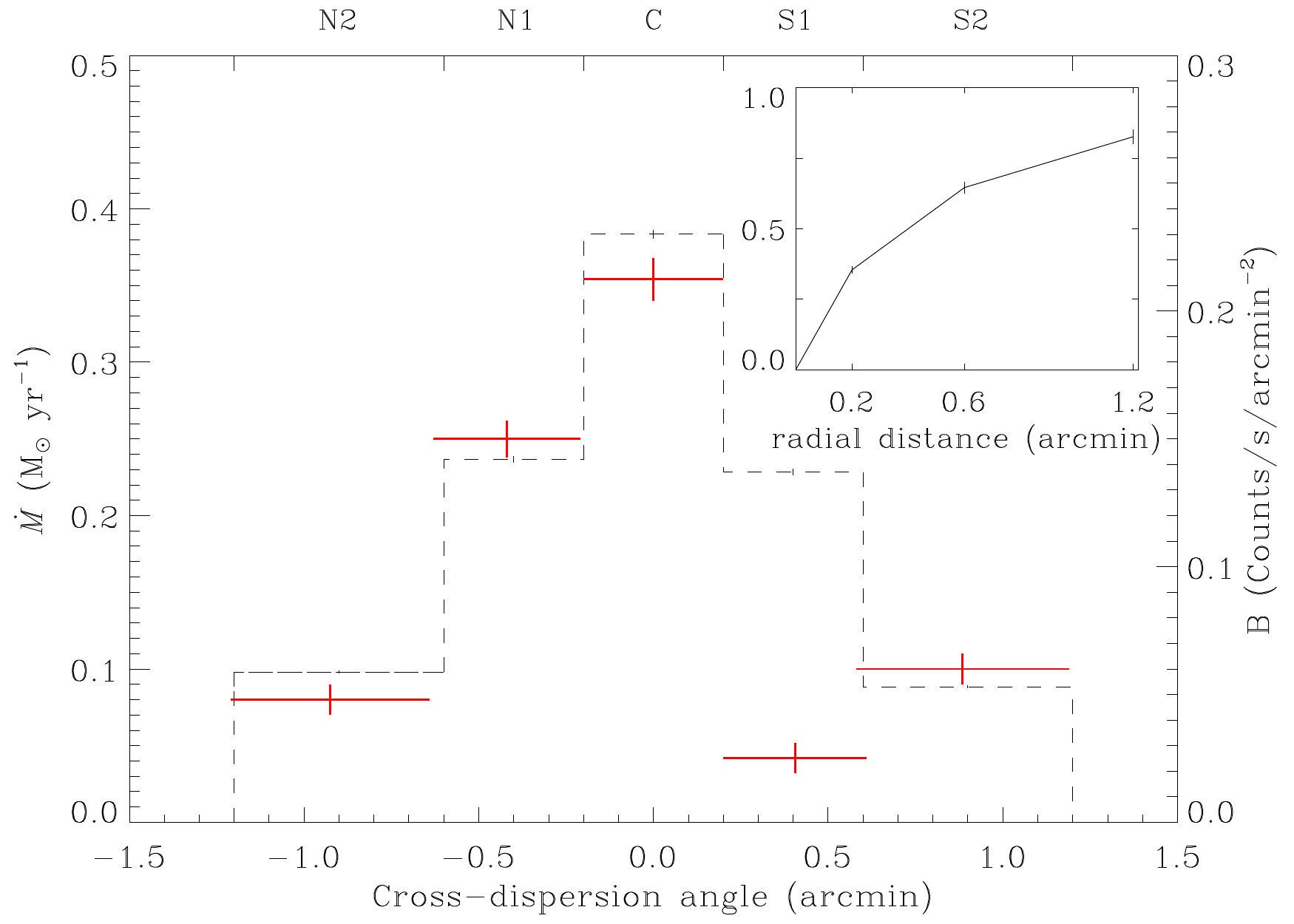}}
\caption{Cooling flow fits (red data points) with a scaled Chandra surface brightness profile of extraction regions $5\arcmin$ wide. The subplot shows the cumulative cooling rate as a function of cross-dispersion radius.}
\label{figure:cool}
\end{figure}

\subsection{Cooling flow measurements}{\label{cf}}

To estimate the presence and magnitude of a cooling flow within different RGS extraction regions, we fit the RGS spectra with a model combining one CIE component and an isobaric cooling flow model. We used the SPEX cooling flow model, which fits the differential emission measure distribution:

\begin{equation}
\label{equation:cool}
D(T)=\frac{5\dot{M} k}{2\mu m_p \Lambda(T)},
\end{equation}
where $\dot{M}$ is the mass deposition rate and $\Lambda$ the cooling function. We modeled the spectra in the temperature range from 0.1 keV to 1.0 keV, divided into 16 bins.

In the fits we excluded the spectral bands containing the \ion{Fe}{xvii} and \ion{O}{vii} resonance lines to avoid the biases caused by the suppression of these lines. The model assumes the standard ICM value of $\mu=0.618$ and $\mbox{Z}_\sun=0.5$ metallicity, which describes the source plasma reasonably well, in particular in the centermost regions of NGC 4636. However, the plasma within the extraction regions is not isobaric and the metallicity is not constant, and therefore the results are to be considered tentative. The deviations from the cooling flow model abundances were compensated for with a CIE model fitted simultaneously to the data.

The best-fit parameters for the cooling flow are plotted in Fig. \ref{figure:cool} on top of the scaled surface brightness profile of the Chandra fits made over the E+W sections. The traditional cooling flow model suggests that the mass deposition rate scales in proportion to the X-ray surface brightness (\citealt{bregman01}). Our results support this scheme partly, although a large deviation is seen in region S1. The symmetrically extracted N1 region yields more than six times the mass deposition rate compared to the S1, in agreement with the results of Chandra analyses that multiphase gas is mainly present in region N1, but not in S1. Taking all the measurement results presented in this paper into account, we note that such drastic differences in temperature phase distributions between the N and S sides would be naturally explained if in situ cooling were taking place in low turbulence
or laminar gas regions at the north side, while the cooling would be dampened by turbulence at south side. However, despite the low  $\dot{M}$ value
in region S1, the value of S2 yields a similar mass deposition rate as region N2 within the errors. 

In the subplot in Fig. \ref{figure:cool} we show the cumulative $\dot{M}$ of the S and N halves as a function of increasing cross-dispersion angle from the center. The total $\dot{M}$ within the examined regions is $0.82\pm0.03$ M$_{\sun}\mbox{yr}^{-1}$ , of which $\sim 4/5$ is contributed by the regions C-N2. In the C region we obtain $\dot{M}=0.35\pm0.01$ M$_{\sun}\mbox{yr}^{-1}$, which is comparable to the \emph{FUSE} measurements based on \ion{O}{vi} doublet emission from the NGC 4636 central galaxy ($0.5\arcmin\times0.5\arcmin$ aperture) of $0.27$ M$_{\sun}\mbox{yr}^{-1}$ (\citealt{bregman05}). The attained total magnitude of the cooling flow agrees with the predictions of the traditional cooling flow models for the NGC 4636 group (\citealt{chen07}).

\subsection{IGM thermodynamics at the core of the NGC 4636 group}{\label{thermod}}

On the basis of the X-ray images of the NGC 4636 group core, the gas dynamics of the N and S side might be expected
to be similar; both sides have been subjected to energetic AGN outbursts and contain shock fronts and large expanding X-ray bubbles that have the potential of driving turbulence into the surrounding gas. However, high-resolution X-ray spectroscopy adds information to this picture, revealing that the two sides are thermodynamically distinct. The spectral maps and 2 CIE component thermal modeling support this finding. By combining the results presented in this paper, a coarse but coherent picture of the thermodynamics of the IGM at the NGC 4636 group core regions is therefore obtained.

On the north side, the relics of an ancient AGN outburst are confined in the NE side, which on average has hotter plasma phases than the NW side. The north side contains multiphase plasma, of which the coolest gas phases are concentrated in the quiescent NW sectors and obey collective gas dynamics. The resonant scattering of \ion{Fe}{xvii} and \ion{O}{vii} ions is strong, indicating laminar gas dynamics within multiple temperature phases. Spectral modeling suggests that a continuous cooling flow occurs in the N sectors, indicating efficient radiative cooling and relatively little gas heating in the corresponding IGM volume. In contrast, the spectral evidence suggests that turbulence dominates the gas dynamics in the south side. The amount of gas in phases of $kT<0.5$ keV is diminished, and the cooling flow is suppressed. The southern side IGM appears to be subjected to turbulent heating and not to radiative cooling. 

We present here three different scenarios that could lead to the disparities observed between the two sides. First, the inclination of the AGN jet axis with respect to the line of sight may be such that the X-ray cavity at the S side induces turbulent velocities in the IGM toward the observer. In the N side, the cavity would then be situated behind much of the gas phases responsible for the resonant scattering, thus increasing the optical depths for a large part of the IGM-emitted photons in our line of sight. 
Our second scenario is that turbulence is present in both sides, but the turbulence properties are anisotropic between the two sides. Namely, in the north side the tangential velocity component of the turbulence field would be large but the radial component suppressed, leading to a high resonant scattering efficiency in this side. To the south, the turbulent field radial component would be larger, suppressing the scattering efficiency in this side.
As our third scenario we propose that sloshing gas flows drive the turbulence observed in the S side. In this case, the RGS extraction regions in the north would be largely collecting photons from the stationary zone, while the southern regions would collect the photons from the zone affected by the turbulent sloshing flow. Therefore radiative cooling would be pronounced in N side, leading to a multiphase gas structure, whereas in the S side the gas would be heated by turbulence mixing and dissipation.

%
\section{Conclusions}{\label{concl}}

Observations of \ion{Fe}{xvii} resonant scattering revealed an asymmetric velocity field distribution in the core regions of the NGC 4636 group. In particular, we observed an asymmetry between the southern and northern parts of the X-ray halo; the analyses suggest that the IGM gas is turbulent in the south side, while the center and north side spectra express characteristics of collective gas motions. 

The X-ray images show that both N and S sides have been subjected to energetic AGN outbursts in the relatively recent past, and spectral maps show evidence of core sloshing with motion toward
the NE. Both of these processes are capable of driving turbulence in the IGM gas. We suggest three alternatives to interpret our observation results: \newline
1) The IGM gas is turbulent in both sides, but the inclination of the AGN jet axis and 3D distribution of turbulent and laminar gas volumes are such that differences in the resonant scattering are observed in our line of sight.
\newline
2) The IGM gas is turbulent in both sides, but the microturbulence velocity distributions are anisotropic between the N and S sides, so that in our line of sight, the radial microturbulence velocity distribution is smaller in the N side than in the S side.
\newline
3) The core-sloshing gas flows drive turbulence to the S side and are the dominating source of microturbulence in the core region, leading to the distinct thermodynamical properties observed in the two sides.

The modeling of cooling the flow yields a cumulative mass deposition rate of $\dot{M}\sim0.8$ M$_{\sun}$yr$^{-1}$ within the $2.4\arcmin\times 5.0\arcmin$ solid angle, a magnitude that generally agrees with the predictions of the traditional cooling flow model for NGC 4636 and \ion{O}{vi} observations made in the FUV band. Nevertheless, we find more than twice the mass deposition rate in the N side than the S side, instead of $\dot{M}$ being proportional to the X-ray surface brightness profile. Overall, we find that the \ion{Fe}{xvii} and \ion{O}{vii} resonant scattering, the magnitude of the cooling flow, and the presence of multiphase gas, including the high concentrations of $kT\lesssim0.4$ keV gas phases, are emphasized in regions C, N1, and N2.

The analysis suggests that concurrent cooling and heating of IGM gas occurs in the core regions of the NGC 4636 galaxy group.

\begin{acknowledgements}
JA and FA wish to acknowledge the Finnish Academy award, decision 266918.

\end{acknowledgements}
\bibliographystyle{aa}
\bibliography{ref}

\end{document}